\documentclass[twocolumn]{aastex7}
\usepackage{amsmath}

\usepackage{placeins}

\usepackage{xcolor}

\begin{document}

\raggedbottom

\title{The Milky Way's Early Disk Was Built in Bursts. I. Evidence from Low-angular-momentum, High-$\alpha$ Stars Scattered Out of the Protodisk by Merger Events}

\author[0000-0003-0795-4854]{Tiancheng Sun}
\affiliation{CAS Key Laboratory of Optical Astronomy, National Astronomical Observatories, Chinese Academy of Sciences, Beijing 100101, China}
\email[show]{suntc@bao.ac.cn}

\author[0000-0002-5818-8769]{Maosheng Xiang}
\affiliation{CAS Key Laboratory of Optical Astronomy, National Astronomical Observatories, Chinese Academy of Sciences, Beijing 100101, China}
\affiliation{Institute for Frontiers in Astronomy and Astrophysics, Beijing Normal University,  Beijing 102206, China}
\email[show]{msxiang@nao.cas.cn}

\author[0000-0002-7642-7583]{Shaolan Bi}
\affiliation{Institute for Frontiers in Astronomy and Astrophysics, Beijing Normal University,  Beijing 102206, China}
\affiliation{School of Physics and Astronomy, Beijing Normal University, Beijing 100875, People’s Republic of China}
\email{bisl@bnu.edu.cn}

\author[0000-0002-2874-2706]{Jifeng Liu}
\affiliation{CAS Key Laboratory of Optical Astronomy, National Astronomical Observatories, Chinese Academy of Sciences, Beijing 100101, China}
\affiliation{School of Astronomy and Space Sciences, University of Chinese Academy of Sciences, Beijing 100049, China}
\affiliation{Institute for Frontiers in Astronomy and Astrophysics, Beijing Normal University,  Beijing 102206, China}
\affiliation{New Cornerstone Science Laboratory, National Astronomical Observatories, Chinese Academy of Sciences, Beijing 100101, China}
\email[show]{jfliu@nao.cas.cn}

\author[0000-0001-9293-131X]{Meng Zhang}
\affiliation{CAS Key Laboratory of Optical Astronomy, National Astronomical Observatories, Chinese Academy of Sciences, Beijing 100101, China}
\email{zhangm@nao.cas.cn}

\author[0000-0002-8337-4117]{Yaqian Wu}
\affiliation{CAS Key Laboratory of Optical Astronomy, National Astronomical Observatories, Chinese Academy of Sciences, Beijing 100101, China}
\email{wuyaqian@nao.cas.cn}



\begin{abstract}


Stars born in the high-$\alpha$ disk can later acquire low angular momentum through dynamical scattering during early merger events. Their age--metallicity distribution thus provides a fossil record of the history of both star formation and dynamical perturbations in the early Milky Way disk. Here we identify a sample of 2,078 such stars from more than two million Gaia DR3 subgiants with robust determinations of ages, kinematics, and abundances. From this sample we reconstruct their intrinsic age--metallicity distribution with a forward modeling approach. Our results confirm that these scattered stars are older and more metal-poor than the bulk of the high-$\alpha$ disk on average. In particular, the joint age--metallicity distribution reveals a robust two-component feature: a broad, dominant component with a peak age of $\sim$11.5~Gyr and ${\rm [Fe/H]}\simeq-0.59$, and an older and more metal-poor component with a peak age of $\sim$12.5~Gyr and ${\rm [Fe/H]}\simeq-0.86$. The presence of this two-component structure suggests that the high-$\alpha$ disk experienced episodic phases of star formation and chemical enrichment. The age distribution of the younger Splash component, compared with that of the high-$\alpha$ disk, provides evidence for a substantial dynamical perturbation at a lookback time of 10.8--11.5~Gyr, possibly associated with an early passage of Gaia--Sausage--Enceladus (GSE). The older Splash component, with a peak age of $\sim$12.5~Gyr, may preserve signatures of an earlier interaction, possibly associated with the proposed Low-energy--Kraken--Heracles (LKH) event.

\end{abstract}



\section{Introduction} 

A remarkable feature of the early evolution of the Milky Way was the rapid formation of the old, high-$\alpha$ disk \citep{2022MNRAS.514..689B,2022arXiv220402989C,2022Natur.603..599X,2025NatAs...9..101X}, followed by successive interactions with infalling dwarf galaxies \citep{2018MNRAS.478..611B,2018Natur.563...85H,2026arXiv260118896M}. These early mergers did not merely bring stars into the stellar halo; they also played a fundamental role in dynamically heating and restructuring the pre-existing disk, leaving long-lived imprints on the present-day Galactic structure. With the advent of the Gaia mission \citep{2016A&A...595A...1G,2018A&A...616A...1G} and large spectroscopic surveys \citep{2012RAA....12..735D,2017AJ....154...94M,2019ApJ...883..107C}, it has become evident that a substantial fraction of stars on halo-like, low-angular-momentum orbits formed in situ within the high-$\alpha$ disk. These stars were subsequently scattered onto highly eccentric orbits by early merger-driven perturbations, giving rise to the so-called in situ halo population \citep{2017ApJ...845..101B} or Splash population \citep{2020MNRAS.497.1603G,2020MNRAS.494.3880B}.

Previous studies have shown that the Splash population is, on average, more metal-poor and older than the thick disk \citep{2018ApJ...863..113H,2019A&A...632A...4D,2019NatAs...3..932G,2020MNRAS.492.3816A,2020MNRAS.494.3880B,2020MNRAS.497.1603G, 2022Natur.603..599X}. However, the reported age distributions are not fully consistent. For instance, \citet{2019NatAs...3..932G} identified a red sequence in the Gaia color--magnitude diagram that corresponds to the in situ halo population now commonly referred to as the Splash. They found that its age distribution broadly follows that of the thick disk during the first $\sim2$ Gyr of Galactic evolution and peaks near 14~Gyr, whereas the thick disk exhibits an extended tail toward younger ages. In contrast, \citet{2020MNRAS.494.3880B} reported a Splash age distribution peaking at approximately 11.5~Gyr, markedly older than the thick-disk population in their sample. Resolving these discrepancies is crucial for establishing how early galaxy mergers shaped the formation and evolution of the thick disk. 

Addressing this challenge necessitates a large stellar sample with accurate ages, metallicities, and wide spatial coverage. Subgiant stars are particularly well-suited for this purpose, as their ages can be precisely determined by locating their positions in the Hertzsprung–Russell diagram. It has been demonstrated that stellar ages of subgiants can be determined to a few percent precision, by using luminosity deduced from the Gaia parallax and atmospheric parameters from spectroscopic surveys \citep{2022Natur.603..599X}. Nevertheless, the LAMOST-Gaia subgiant star sample of \citet{2022Natur.603..599X} is mainly confined toward the Galactic anti-center direction, while lacking complete coverage toward the inner disk.

The third Gaia data release \citep[Gaia DR3;][]{2023A&A...674A...1G} provides low-resolution ($R\simeq50$) BP/RP spectra with precise flux calibration for bright stars ($G\lesssim17.5$) across the full sky \citep{2023A&A...674A..27A}. In this work, we identify the Splash population from a sample of 2.16 million subgiant stars with precise age estimates, derived from Gaia DR3 BP/RP spectra. This subgiant star sample not only represents a tenfold increase in sample size relative to that of \cite{2022Natur.603..599X}, but also exhibits much more complete sky coverage. Furthermore, we investigate the age--metallicity distribution of the Splash population and assess the impact of early merger events on the Galactic disk.




This paper is organized as follows. Section~\ref{sec:data} describes the stellar measurements and the construction of the subgiant, high-$\alpha$ disk, and Splash samples. Section~\ref{sec:selection_correction} presents the correction for the subgiant selection function. Section~\ref{sec:results} presents the age--metallicity distributions and the forward modeling of the intrinsic Splash distribution. Section~\ref{sec:results-sec3} discusses the implications for the early star formation and merger history of the Milky Way disk. Finally, Section~\ref{sec:conclusions} summarizes our main conclusions.

\section{Data} \label{sec:data}

\subsection{Stellar Parameters and Orbital Properties}


\begin{figure*}[t]
\plotone{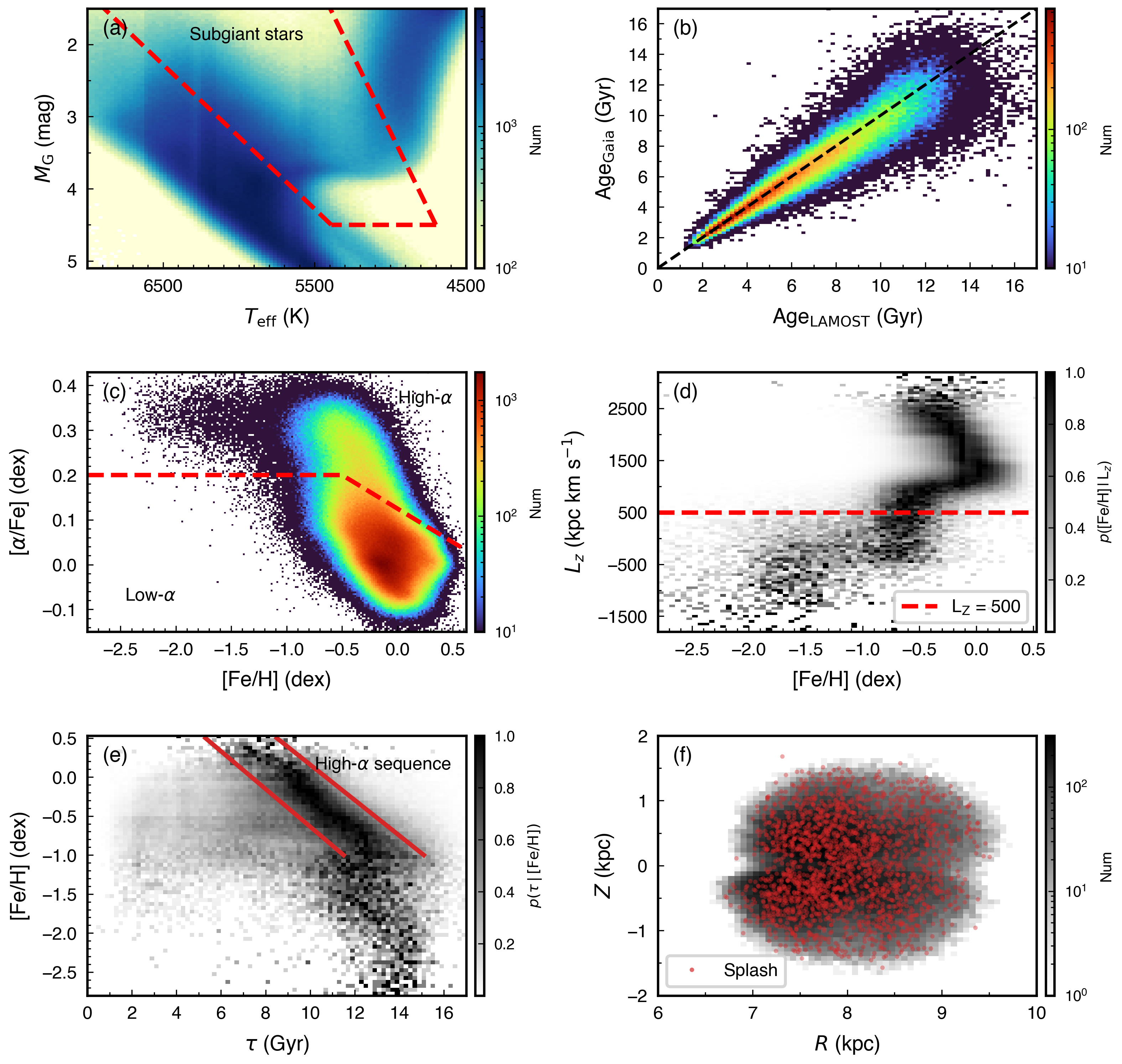}
\caption{(a) Illustration of the subgiant selection in the $T_{\rm eff}$--$M_G$ diagram for stars with metallicities in the range $-0.1<[{\rm Fe/H}]<0.1$. The two red dashed lines show the metallicity-dependent selection boundaries evaluated at solar metallicity, while the horizontal red dashed line marks the additional cut at $M_G=4.5$. Selected subgiants lie between the two sloping boundaries and satisfy $M_G<4.5$. (b) Comparison of age and [Fe/H] estimates for $\sim110,000$ stars in common between the Gaia DR3 subgiant sample and the LAMOST DR7 subgiant sample of \citet{2022Natur.603..599X}. The black line represents the 1:1 line. (c) Number density distribution of sample stars in [Fe/H]--[$\alpha$/Fe] space. The red dashed line separates the sample into high-$\alpha$ and low-$\alpha$ regimes. (d) The row-normalized stellar density of sample stars in [Fe/H]--$L_z$ space. The horizontal dashed line marks $L_z$ = 500 kpc km s$^{-1}$. (e) Probability distribution of stellar age $p(\tau\mid[\mathrm{Fe/H}])$, normalized to the peak value for each [Fe/H], for high-$\alpha$ stars with $L_z$ $<$ 1500 kpc km s$^{-1}$ in the Gaia DR3 subgiant sample. The red boundaries delineate the parameter window in age--[Fe/H] space where the high-$\alpha$ disk is expected to dominate. (f) Spatial distributions, in Galactocentric cylindrical coordinates $R$--$Z$, of the high-$\alpha$ disk (gray-scale density map) and Splash (red points) populations selected from panel (e).
\label{fig0}}
\end{figure*}


Gaia DR3 \citep{2023A&A...674A...1G} provides a catalog of 220 million low-resolution ($R\simeq20-90$) spectra obtained with the slitless spectrophotometers BP (330--680~nm) and RP (640--1050~nm) \citep{2021A&A...652A..86C,2023A&A...674A...3M,2023A&A...674A..27A}. We derived stellar atmospheric parameters ($T_{\rm eff}$, $\log g$, [Fe/H]), elemental abundances, interstellar extinction ($E_{B-V}$), and distance from a combination of the Gaia BP/RP spectra (hereafter ``XP spectra''), photometry ($G$, $BP$, $RP$) and parallax using a tailored version of the \texttt{DD-Payne} \citep{2019ApJS..245...34X,2025ApJS..279....5Z}, a hybrid method that couples data-driven modeling with physically informed priors. \texttt{DD-Payne} is a forward modeling method that can simultaneously infer stellar parameters and their covariance. Systematic errors in the Gaia parallax measurements are corrected following the method of \citet{2021A&A...649A...2L}. To construct the \texttt{DD-Payne} models, we adopted stars with high-fidelity labels from LAMOST DR9 \citep{2025ApJS..279....5Z} as the training sample; these models are then employed to infer stellar parameters for the remaining Gaia stars. A detailed introduction to the stellar parameter determination will be presented elsewhere.

The individual abundance estimates suffer substantial uncertainties ($\gtrsim0.1$~dex) due to the limited spectral resolution. To obtain more robust $\alpha$-elemental abundances, we further estimate an overall $\alpha$-to-iron abundance ratio [$\alpha$/Fe] with a data-driven approach, by mapping the Gaia XP spectra and the \texttt{DD-Payne} labels (atmospheric parameters and abundances), as well as the orbit eccentricity and actions ($J_R$, $J_\phi$, $J_z$) to the [$\alpha$/Fe] using the XGBoost algorithm \citep{2016arXiv160302754C}. Additionally, we derived a data-driven version of the absolute magnitude ($M_G$) from the Gaia XP spectra, photometry, and parallax using the XGBoost algorithm.
%

Using radial velocities and proper motions from Gaia DR3, we compute the angular momentum $L_z$ and orbital eccentricity of our sample stars with Galpy \citep{2015ApJS..216...29B}, adopting the MWPotential2014 potential. We assume the Sun is located at $R_{\odot}$ = 8.178 kpc \citep{2019A&A...625L..10G} and $Z_{\odot}$ = 10 pc above the Galactic midplane \citep{2018ApJS..237...33X}, with a local standard of rest (LSR) velocity of 220 km s$^{\rm -1}$. The solar motion relative to the LSR is taken as $(U_\odot, V_\odot, W_\odot) = (-7.01, 10.13, 4.95)\,\mathrm{km\,s^{-1}}$ \citep{2015MNRAS.449..162H}.

\subsection{Subgiant Selection and Age Determination}


\begin{table}[t]
\centering
\caption{Slope and intercept of the linear functions for the upper and lower boundaries of the subgiant star sample selection.}
\label{tab:subgiant_selection}
\begin{tabular}{ccccc}
\hline
\hline
[Fe/H] & slope1 & zpt1 & slope2 & zpt2 \\
\hline
0.4  & $-0.0015$ & 10.82 & $-0.0019$ & 14.37 \\
0.2  & $-0.0029$ & 17.72 & $-0.0020$ & 14.94 \\
0.0  & $-0.0043$ & 24.71 & $-0.0020$ & 15.28 \\
$-$0.2 & $-0.0044$ & 25.50 & $-0.0020$ & 15.21 \\
$-$0.4 & $-0.0039$ & 23.42 & $-0.0020$ & 15.25 \\
$-$0.6 & $-0.0039$ & 23.59 & $-0.0020$ & 15.43 \\
$-$0.8 & $-0.0034$ & 21.99 & $-0.0018$ & 14.51 \\
$-$1.0 & $-0.0036$ & 23.20 & $-0.0018$ & 14.31 \\
$-$1.2 & $-0.0042$ & 25.93 & $-0.0016$ & 13.56 \\
$-$1.4 & $-0.0032$ & 20.96 & $-0.0015$ & 12.88 \\
$-$1.6 & $-0.0038$ & 24.33 & $-0.0016$ & 13.50 \\
$-$1.8 & $-0.0046$ & 28.67 & $-0.0016$ & 13.77 \\
$-$2.0 & $-0.0043$ & 27.68 & $-0.0014$ & 12.73 \\
$-$2.2 & $-0.0047$ & 30.05 & $-0.0016$ & 13.97 \\
$-$2.65 & $-0.0048$ & 30.30 & $-0.0016$ & 14.19 \\
\hline
\end{tabular}

\vspace{0.5em}
\footnotesize\raggedright
The boundary of subgiant stars in the $T_{\rm eff}$–$M_G$ diagram is given by $M_G = {\rm slope} \times T_{\rm eff} + {\rm zpt}$. The slopes and intercepts (‘zpt’) listed in the table are adopted as anchors for interpolation to match the measured [Fe/H] of each star. In addition to the two metallicity-dependent linear boundaries, all selected subgiants are required to satisfy $M_G<4.5$.
\end{table}

Subgiant stars were identified in the $T_{\rm eff}$--$M_G$ diagram, as illustrated in Figure~\ref{fig0}. Because the subgiant locus and the morphology of the evolutionary sequences vary with [Fe/H], we evaluated the selection boundaries at the metallicity grid points listed in Table~\ref{tab:subgiant_selection}. At each metallicity grid point, we used Yale--Yonsei (YY) isochrones \citep{2004ApJS..155..667D} to locate the main-sequence turnoff and the base of the red-giant branch, which together delineate the expected subgiant region. The boundaries separating subgiants from dwarfs and red giants were then parameterized as linear relations in the $T_{\rm eff}$--$M_G$ plane; the corresponding slopes and intercepts are listed in Table~\ref{tab:subgiant_selection}. The subgiant sample is defined by stars lying between these two metallicity-dependent relations and satisfying the additional absolute-magnitude cut $M_G<4.5$.


The selection boundaries for stars of solar metallicity are shown as an example in Figure~\ref{fig0}(a). To ensure that the boundaries vary smoothly with [Fe/H], we interpolated the tabulated slopes and intercepts to the measured metallicity of each star. This procedure yields a catalog of approximately $8.7$ million subgiants, including $\sim3.2$ million with radial velocities from Gaia RVS spectra \citep{2023A&A...674A...5K} and $\sim5.5$ million without radial-velocity measurements. In this work, we use only stars with Gaia RVS radial velocities.

The ages of the subgiant stars were determined by matching their stellar parameters ($T_{\rm eff}$, $[\mathrm{Fe/H}]$, and $[\alpha/\mathrm{Fe}]$) and absolute magnitudes against the YY isochrones \citep{2004ApJS..155..667D} using a Bayesian approach. From the full posterior distributions of the age estimates, we computed the posterior-weighted mean and standard deviation, adopting the latter as the age uncertainty. We also examined the mode, median, and the 16th and 84th percentile ages. Since our subgiant sample consists of bright, nearby stars, their Gaia parallaxes provide strong constraints on their luminosities, leading to only minor asymmetries in the posterior distributions. Consequently, the mean ages are in good agreement with the mode and median ages in the vast majority of cases.

To ensure the reliability of atmospheric parameters and age estimates, we retained only stars with effective temperature uncertainties below 150~K and metallicity uncertainties below 0.15~dex. We also applied a quality cut based on the Gaia re-normalized unit weight error \citep[RUWE;][]{2021A&A...649A...2L}, requiring $\mathrm{RUWE} < 1.2$. 
We further restricted the sample to stars with $T_{\rm eff}$ $<$ 6800 K, where DD-Payne abundances are most reliable, and $\log g > 3$ to exclude contaminants such as He-burning horizontal branch stars. Stars whose $2\sigma$ lower age limits exceeded the age of the Universe \citep[13.8~Gyr;][]{2016A&A...594A..13P} were also removed.\footnote{Specifically, we removed stars satisfying $\mathrm{age}-2\sigma_{\rm age}>13.8$~Gyr.}
Finally, we restricted the sample to stars with relative age precision better than 15\%, resulting in a final sample of $\sim2.16$ million stars with a median age uncertainty of 6.7\%.

Figure \ref{fig0}(b) presents a comparison between our subgiant sample and the common sources in \citet{2022Natur.603..599X} ($\sim$110,000 stars). Overall, stellar ages show good agreement between the two datasets. However, for stars older than 8~Gyr, our inferred ages are systematically slightly younger than those reported by \citet{2022Natur.603..599X}.
The offset likely arises from methodological differences: our DD-Payne measurements follow the updated implementation of \citet{2025ApJS..279....5Z}, whereas \citet{2022Natur.603..599X} employed an earlier version \citep{2019ApJS..245...34X}, which produces small differences in [Fe/H] estimates that propagate into age determinations.


\subsection{Identification of the High-\texorpdfstring{$\alpha$}{alpha} Disk and Splash Populations}

To identify Splash stars, we first separate the high-$\alpha$ population from the low-$\alpha$ population using an empirical boundary in the [Fe/H]--[$\alpha$/Fe] plane (Figure~\ref{fig0}(c)). We then identify the high-$\alpha$-disk sequence in the age--[Fe/H] plane using two empirical boundaries that enclose its high-density locus, where the stellar number density is normalized along the age direction at fixed [Fe/H] (Figure~\ref{fig0}(e)). Specifically, we define our high-$\alpha$ disk sample by requiring
\begin{linenomath*}
\[
\left\{
\begin{aligned}
& [\alpha/\mathrm{Fe}] >
  \begin{cases}
    0.20, & [\mathrm{Fe/H}] < -0.5, \\
    -0.15[\mathrm{Fe/H}] + 0.125, & [\mathrm{Fe/H}] \geq -0.5,
  \end{cases} \\
& -0.242\,\tau + 1.782 < [\mathrm{Fe/H}]
  < -0.227\,\tau + 2.432, \\
& [\mathrm{Fe/H}] > -1, \\
& 500 < L_z < 1500\ \mathrm{kpc\,km\,s^{-1}},
\end{aligned}
\right.
\]
\end{linenomath*}
where $\tau$ is the stellar age in Gyr. These criteria yield a high-$\alpha$ disk sample of 145,410 stars.

Splash stars were selected by retaining the same chemical, age--metallicity, and metallicity criteria as above, while replacing the angular-momentum range with $L_z < 500~\mathrm{kpc\,km\,s^{-1}}$. Following \citet{2022Natur.603..599X}, this criterion identifies stars with chemical and age--metallicity properties characteristic of the high-$\alpha$ disk but substantially lower angular momenta. It selects the downward-extending feature in Figure~\ref{fig0}(d), along which $L_z$ shows little further variation toward lower metallicity. The resulting sample contains 2,078 candidate Splash stars, whose distribution in the $R$--$Z$ plane is shown in Figure~\ref{fig0}(f).

\section{Correction for the Subgiant Selection Function}

\label{sec:selection_correction}

Having defined the high-$\alpha$ disk and Splash samples, we correct for the selection effects introduced by the construction of the parent subgiant catalog. The metallicity-dependent boundaries in the $T_{\rm eff}$--$M_G$ plane select a restricted segment of each isochrone. Although the parent subgiant sample additionally requires $M_G<4.5$, this magnitude cut does not further restrict the isochrone segments used to compute the selection function, because all relevant isochrone points already satisfy this criterion. The fraction of the underlying stellar population represented by the selected segment depends on age and chemical composition, leading to differential sampling that can bias the inferred age and metallicity distributions and their joint structure. To account for this effect, we assign each star a subgiant selection factor, $s_i=\texttt{MASSRATIOINI}(\tau_i,[\mathrm{Fe/H}]_i,[\alpha/\mathrm{Fe}]_i)$. Here, $\texttt{MASSRATIOINI}$ denotes the fraction of the total initial stellar mass represented by the portion of the corresponding isochrone lying within the metallicity-dependent subgiant-selection boundaries, evaluated using the \citet{2001MNRAS.322..231K} initial mass function. The selection function is tabulated on a regular grid spanning $5.5\leq\tau\leq15.5$~Gyr in steps of 0.5~Gyr, $-1.0\leq[\mathrm{Fe/H}]\leq0.4$ in steps of 0.2~dex, and $[\alpha/\mathrm{Fe}]=0.0$, 0.2, and 0.4~dex. We evaluate $s_i$ by trilinear interpolation within the grid and linear extrapolation beyond its boundaries. None of the 2,078 Splash stars require extrapolation in age or [Fe/H], and only 15 require extrapolation in $[\alpha/\mathrm{Fe}]$. Each star is assigned a raw inverse-selection weight, $q_i=s_i^{-1}$. For the Splash sample, these weights have a median of 214.3 and a 16th--84th percentile range of 200.7--246.2. We normalize the weights separately for each sample to obtain unit-mean weights, $\omega_i=q_i/\overline{q}$, as defined in Equation~\ref{eq:selection_weights}, where $\overline{q}$ is the mean raw weight of the corresponding sample. For the Splash sample, $\overline{q}\simeq224.4$, giving normalized weights with a median of 0.955 and a 16th--84th percentile range of 0.894--1.097. These normalized weights are incorporated consistently into the distribution estimates, likelihood fits, bootstrap realizations, and goodness-of-fit diagnostics described below. Figure~\ref{app-selection} in the Appendix compares the age and metallicity distributions of the Splash and high-$\alpha$ disk samples before and after the selection correction.

\section{Results} \label{sec:results}

\begin{figure*}[t]
\plotone{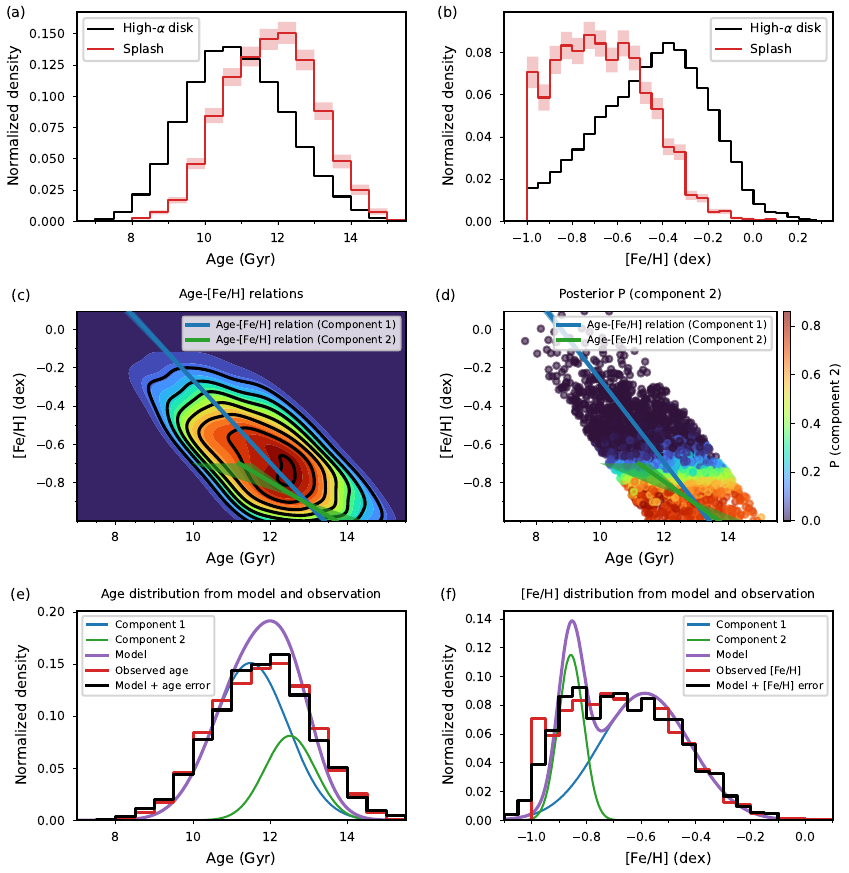}
\caption{Two-Gaussian decomposition of the age--metallicity ($[\mathrm{Fe/H}]$) distribution of the Splash population. (a, b) Selection-corrected age and metallicity distributions, respectively, of the high-$\alpha$ disk (black) and Splash population (red). Red shading shows the selection-weighted counting uncertainty of the Splash population, calculated in each bin $B$ as $\sigma_{W_B}=(\sum_{i\in B}\omega_i^2)^{1/2}$ and normalized by the total selection weight. (c) Inverse-selection-weighted KDE of Splash stars in the age--$[\mathrm{Fe/H}]$ plane; colors and black contours indicate the smoothed density and its iso-density levels. Blue and green lines show the best-fitting relations of the two Gaussian components, with shaded pointwise 16th--84th percentile intervals derived from bootstrap refits. 
The green relation and its uncertainty band are displayed only in the metal-poor regime, $[\mathrm{Fe/H}] \leq -0.7$, where Component~2 is concentrated. (d) Posterior probability that each star belongs to Component~2, with the same fitted relations and uncertainty bands as in panel (c). (e, f) Comparison between the selection-corrected observed age and $[\mathrm{Fe/H}]$ distributions and the predictions of the best-fitting intrinsic model, respectively. Blue and green curves show the individual model components, the purple curve their sum, and red and black histograms show the selection-corrected observations and mock realizations drawn from the best-fitting model, respectively. Mock ages include a 6.7\% fractional uncertainty, while mock metallicity uncertainties are drawn locally from observed stars of similar [Fe/H].
\label{fig1}}
\end{figure*}

\subsection{Age and metallicity distributions}

Figure~\ref{fig1}(a--b) presents the selection-corrected age and metallicity distributions of the Splash stars. For comparison, we also show the corresponding distributions of the high-$\alpha$ disk. The Splash age distribution peaks at $\sim12.25$ Gyr, substantially earlier than the high-$\alpha$ disk peak at $\sim10.75$ Gyr, indicating that the Splash is systematically older. The Splash is also markedly more metal-poor than the high-$\alpha$ disk. In particular, its broad metallicity distribution suggests that a single Gaussian may not provide an adequate description, motivating the forward modeling of the intrinsic joint age--$[\mathrm{Fe/H}]$ distribution in Section~\ref{sec:results-sec1}.

The corrected distributions were constructed using normalized
inverse-selection weights,
\begin{linenomath*}
\begin{equation}
\begin{aligned}
q_i &= s_i^{-1},\\
\omega_i
&= \frac{q_i}{\langle q\rangle}
 = \frac{Nq_i}{\displaystyle\sum_{j=1}^{N}q_j},
\end{aligned}
\label{eq:selection_weights}
\end{equation}
\end{linenomath*}
where $\langle q\rangle=N^{-1}\sum_{j=1}^{N}q_j$. The weights were normalized separately for the Splash and high-$\alpha$ disk samples;
by construction, $\sum_{i=1}^{N}\omega_i=N$ in each case. For a histogram bin $B$, the normalized bin value and its
selection-weighted counting uncertainty are
\begin{linenomath*}
\begin{equation}
\begin{aligned}
P_B &=
\frac{\displaystyle\sum_{i\in B}\omega_i}
     {\displaystyle\sum_{i=1}^{N}\omega_i},\\[3pt]
\sigma_{P_B} &=
\frac{\left(\displaystyle\sum_{i\in B}\omega_i^2\right)^{1/2}}
     {\displaystyle\sum_{i=1}^{N}\omega_i}.
\end{aligned}
\label{eq:weighted_histogram}
\end{equation}
\end{linenomath*}
The shaded regions associated with the Splash distributions in Figure~\ref{fig1}(a--b) show the corresponding uncertainties.

\subsection{Forward modeling of the age--[Fe/H] distribution} \label{sec:results-sec1}

Although the distributions presented above have been corrected for differential subgiant selection, they remain affected by the measurement errors. To infer the intrinsic joint age--$[\mathrm{Fe/H}]$ distribution of the Splash population, we model the observed stars with a weighted Gaussian mixture model that explicitly incorporates the individual measurement uncertainties in both quantities. Our model contains two components. For each component $k$, we assume that the latent stellar metallicity follows a Gaussian distribution,
\begin{linenomath*}
\begin{equation}
P([\mathrm{Fe/H}]) \sim
\mathcal{N}
\left(
\mu_{{\rm [Fe/H]},k},
\delta_{{\rm [Fe/H]},k}^{2}
\right),
\end{equation}
\end{linenomath*}
while the stellar age at fixed metallicity is described by a linear relation with intrinsic scatter,
\begin{linenomath*}
\begin{equation}
P(\tau\mid[\mathrm{Fe/H}]) \sim
\mathcal{N}
\left(
a_k[\mathrm{Fe/H}]+b_k,
\delta_{\tau,k}^{2}
\right).
\end{equation}
\end{linenomath*}

Using the normalized inverse-selection weights defined in
Equation~\ref{eq:selection_weights}, the selection-corrected weighted
log-likelihood is
\begin{linenomath*}
\begin{equation}
\ln\mathcal{L}_{\rm w}
=
\sum_i \omega_i
\ln\left[
\sum_k w_k\,
\mathcal{N}\left(
\boldsymbol{x}_i
\mid
\boldsymbol{\mu}_k,
\boldsymbol{C}_k+\boldsymbol{C}_{{\rm obs},i}
\right)
\right],
\label{eq:weighted_likelihood}
\end{equation}
\end{linenomath*}
where $\boldsymbol{x}_i=(\tau_i,[\mathrm{Fe/H}]_i)$ is the observed
age--metallicity vector of star $i$, $w_k$ is the mixture fraction of
component $k$, and $\omega_i$ is its normalized inverse-selection
weight.
For this parameterization, the mean vector of component k is
\begin{linenomath*}
\begin{equation}
\boldsymbol{\mu}_k =
\begin{pmatrix}
a_k\mu_{{\rm [Fe/H]},k} + b_k \\
\mu_{{\rm [Fe/H]},k}
\end{pmatrix}.
\end{equation}
\end{linenomath*}
The corresponding intrinsic covariance matrix is
\begin{linenomath*}
\begin{equation}
\mathbf{C}_k =
\begin{pmatrix}
a_k^2\delta_{{\rm [Fe/H]},k}^2 + \delta_{\tau,k}^2
&
a_k\delta_{{\rm [Fe/H]},k}^2
\\
a_k\delta_{{\rm [Fe/H]},k}^2
&
\delta_{{\rm [Fe/H]},k}^2
\end{pmatrix}.
\end{equation}
\end{linenomath*}
The observational covariance matrix for star $i$ is approximated as diagonal:
\begin{linenomath*}
\begin{equation}
\mathbf{C}_{{\rm obs},i} =
\begin{pmatrix}
\sigma_{\tau,i}^2 & 0 \\
0 & \sigma_{{\rm [Fe/H]},i}^2
\end{pmatrix},
\end{equation}
\end{linenomath*}
where $\sigma_{\tau,i}$ and $\sigma_{{\rm [Fe/H]},i}$ are the reported marginal uncertainties in age and metallicity, respectively. Because $[\mathrm{Fe/H}]$ is an input to the isochrone-based age determination, the two errors may be correlated. In the absence of star-by-star estimates of this covariance, we adopt the diagonal form as a practical approximation. 
The best-fitting parameters are obtained by maximizing the selection-corrected weighted log-likelihood given by Equation \ref{eq:weighted_likelihood}.

As illustrated in Figure~\ref{fig1}(c--f), the best-fitting two-component model separates the Splash population into a dominant, relatively metal-rich component and a secondary, older and more metal-poor component. From 500 bootstrap resamplings, we obtained asymmetric 16th--84th percentile intervals relative to the full-sample best fit; in each realization, stars were resampled with replacement together with their inverse-selection weights. The resampled weights were renormalized to have a mean value of unity, and the selection-corrected weighted likelihood was then refitted. The two components were reordered by increasing derived mean age, $\mu_\tau=a\mu_{\rm [Fe/H]}+b$, to avoid label switching.

Component~1 has a selection-corrected mixture fraction of $0.73_{-0.03}^{+0.04}$ and a metallicity centroid of $\mu_{\rm [Fe/H]}=-0.59_{-0.01}^{+0.01}$~dex, with an intrinsic metallicity dispersion of $\delta_{\rm [Fe/H]}=0.165_{-0.003}^{+0.003}$~dex. Its
age--metallicity relation is described by $\tau=a[\mathrm{Fe/H}]+b$, with $a=-4.61_{-0.24}^{+0.20}~{\rm Gyr~dex^{-1}}$ and $b=8.78_{-0.13}^{+0.10}$~Gyr. These parameters correspond to a derived mean age of $\mu_\tau=11.49_{-0.05}^{+0.05}$~Gyr and a marginal age dispersion of $\sigma_{\tau,\mathrm{marg}}=0.97_{-0.03}^{+0.03}$~Gyr.

Component~2 accounts for the remaining selection-corrected mixture fraction of $0.27_{-0.04}^{+0.03}$ and is centered at lower
metallicity and older ages. It has $\mu_{\rm [Fe/H]}=-0.86_{-0.01}^{+0.01}$~dex, $\delta_{\rm [Fe/H]}=0.046_{-0.014}^{+0.009}$~dex,
$a=-8.37_{-6.48}^{+1.65}~{\rm Gyr~dex^{-1}}$, and $b=5.34_{-5.79}^{+1.48}$~Gyr. The corresponding derived mean age and marginal age dispersion are $\mu_\tau=12.51_{-0.09}^{+0.06}$~Gyr and $\sigma_{\tau,\mathrm{marg}}=0.66_{-0.06}^{+0.04}$~Gyr, respectively. Thus, the two-component model identifies a dominant Splash component with a broader metallicity distribution and a secondary component that is older, more metal-poor, and more compact in both metallicity and marginal age. The metal-poor component has a steeper best-fitting slope, although its slope and intercept remain poorly constrained by the bootstrap realizations.

\begin{figure*}[t]
\centering
\includegraphics[width=1\textwidth]{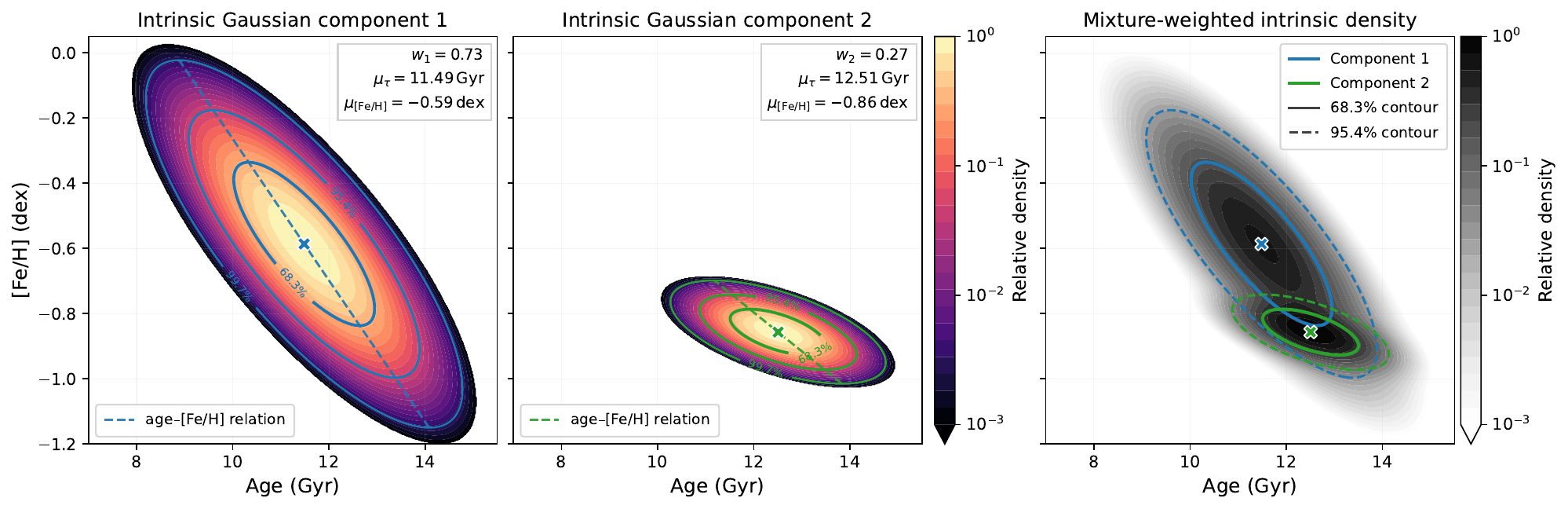}
\caption{Intrinsic age--$[\mathrm{Fe/H}]$ density distributions of the two Gaussian components inferred for the Splash population. The left and middle panels show the two components separately, evaluated from their intrinsic covariance matrices before convolution with the individual age and metallicity uncertainties. The color shading shows the density normalized by the peak of each component on a logarithmic scale, and the contours enclose 68.3\%, 95.4\%, and 99.7\% of the corresponding bivariate Gaussian probability. The dashed lines show the fitted conditional mean age at each metallicity, $\tau=a[\mathrm{Fe/H}]+b$. Because the model allows an intrinsic spread in age at fixed $[\mathrm{Fe/H}]$, these lines do not necessarily coincide with the major axes of the two-dimensional Gaussian contours. The crosses mark the component means, while the mixture weights and mean ages and metallicities are listed in the upper-right corners. The right panel shows the mixture-weighted intrinsic density, $p_{\rm mix}=w_1p_1+w_2p_2$, normalized by its peak. The solid and dashed blue and green curves indicate the 68.3\% and 95.4\% contours of the individual components, respectively.
\label{fig1-gauss-2d}}
\end{figure*}

These two fitted Gaussian components are visualized in the joint age--$[\mathrm{Fe/H}]$ plane in Figure~\ref{fig1-gauss-2d}. Their distinct distributions indicate that the Splash is not a homogeneous population in age and metallicity. The older, more metal-poor component may trace an earlier phase of star formation and chemical enrichment, whereas the dominant, relatively younger and more metal-rich component may represent a subsequent phase. The two-component structure therefore suggests that the Splash population preserves evidence of a complex and potentially episodic formation history.

\begin{table*}[t]
\centering
\caption{Model comparison for the intrinsic age--metallicity distribution of Splash stars with $[\mathrm{Fe/H}]>-1$. Lower AIC and BIC values indicate a preferred model after penalizing model complexity. The KS statistics compare the selection-corrected, inverse-selection-weighted empirical distributions with error-convolved mock distributions drawn from the fitted intrinsic models. The corresponding $p$-values are calibrated using 5,000 selection-aware parametric-bootstrap realizations for each model rather than the standard unweighted KS null distribution. Smaller $D$ values and larger $p$-values indicate better agreement between the model and the data.}
\label{tab:model-comparison}
\begin{tabular}{lcccc}
\hline
Model & AIC & BIC & Age KS test $(D,p)$ & $[\mathrm{Fe/H}]$ KS test $(D,p)$ \\
\hline
Two-component Gaussian mixture & 4825.17 & 4887.20 & 0.0168, 0.9466 & 0.0202, 0.8296  \\
Single-Gaussian & 4955.46 & 4983.66 & 0.0526, 0.0160 & 0.0722, 0.0006 \\
Gaussian plus power law & 4854.08 & 4921.75 & 0.0411, 0.0918 & 0.0387, 0.1170 \\
\hline
\end{tabular}
\end{table*}

These two components provide a good reconstruction of the observed data after observational errors are incorporated (Figure~\ref{fig1}(e--f)). The model reproduces both the observed age distribution (Figure~\ref{fig1}(a)) and the broad structure of the $[\mathrm{Fe/H}]$ distribution (Figure~\ref{fig1}(b)). Quantitatively, the two-Gaussian model yields KS statistics of 0.0168 for age and 0.0202 for $[\mathrm{Fe/H}]$, with corresponding $p$-values of 0.9466 and 0.8296, respectively, indicating good agreement with the observed distributions.

\subsection{Statistical Model Assessment and Robustness Tests}

To assess whether the two-Gaussian model is statistically required relative to simpler or alternative functional forms, we compare the models using both information criteria and goodness-of-fit tests. Specifically, we compute the Akaike Information Criterion (AIC) and the Bayesian Information Criterion (BIC), defined as
\begin{linenomath*}
\begin{equation}
\mathrm{AIC}
=
2k-2\ln\mathcal{L}_{{\rm w},\max},
\end{equation}
\end{linenomath*}
and
\begin{linenomath*}
\begin{equation}
\mathrm{BIC}
=
k\ln N-2\ln\mathcal{L}_{{\rm w},\max}.
\end{equation}
\end{linenomath*}
where $k$ is the number of free parameters, $\mathcal{L}_{{\rm w},\max}$ is the maximum selection-corrected weighted likelihood, and $N=2078$ is the number of observed Splash stars. Under the unit-mean normalization adopted in Equation~\ref{eq:selection_weights}, $\sum_i\omega_i=N$.
Since the inverse-selection weights are incorporated as replication weights, $\mathcal{L}_{{\rm w},\max}$ is formally a weighted pseudo-likelihood. The resulting AIC and BIC values are therefore used for relative comparisons among models fitted to the same sample
with the same normalized weights.
In both cases, smaller values indicate a preferred model, with the BIC imposing a stronger penalty on model complexity, especially for large samples. 


In addition, we assess the marginal goodness of fit using a selection-corrected weighted Kolmogorov--Smirnov (KS) statistic. For each marginal distribution, the empirical cumulative distribution function of the observed sample is constructed using the normalized inverse-selection weights, whereas the error-convolved mock sample drawn from the fitted intrinsic model is treated as unweighted. Because the standard KS null distribution does not apply to this weighted statistic, the corresponding $p$-values are calibrated using selection-aware parametric-bootstrap realizations that explicitly reproduce the selection process. Because the intrinsic mixture model is defined only in age and [Fe/H], an $[\alpha/\mathrm{Fe}]$ value is additionally assigned to each simulated star before evaluating the three-dimensional subgiant selection function. Specifically, for each simulated star, we draw $[\alpha/\mathrm{Fe}]$ from observed Splash stars within 0.2~dex in [Fe/H], with drawing probabilities proportional to their inverse-selection weights. When fewer than 100 stars fall within this interval, the 100 nearest Splash stars in [Fe/H] are used instead. This procedure approximately samples the selection-corrected empirical $p([\alpha/\mathrm{Fe}]\mid[\mathrm{Fe/H}])$ distribution. The resulting age, [Fe/H], and $[\alpha/\mathrm{Fe}]$ values are used to evaluate the same selection function applied to the data. In each realization, a pseudo-observed sample is drawn according to these selection probabilities and corrected using the corresponding inverse-selection weights, and its weighted KS statistic is then computed relative to an independent error-convolved realization of the fitted model. We use 5,000 parametric-bootstrap realizations for each model, with the fitted model parameters held fixed during the calibration. The resulting information criteria and KS statistics are summarized in Table~\ref{tab:model-comparison}.

The single-Gaussian model is strongly disfavored relative to the best-fitting two-Gaussian model. The two-Gaussian model yields
$\mathrm{AIC}=4825.17$ and $\mathrm{BIC}=4887.20$, whereas the single-Gaussian model gives $\mathrm{AIC}=4955.46$ and $\mathrm{BIC}=4983.66$. The corresponding differences, $\Delta\mathrm{AIC}=130.29$ and $\Delta\mathrm{BIC}=96.46$, provide strong statistical evidence in favor of the two-Gaussian description. The marginal KS tests lead to the same conclusion. For the age distribution, the single-Gaussian model gives $D=0.0526$ and $p=0.0160$, while for the $[\mathrm{Fe/H}]$ distribution it gives $D=0.0722$ and $p=0.0006$. Although the KS statistics are numerically small, the low $p$-values indicate statistically significant departures from a single-Gaussian description in both marginal distributions, with a slightly stronger discrepancy in $[\mathrm{Fe/H}]$ (see Appendix Figure~\ref{fig1-app-splash}(e--f)). Together, these results indicate that the structure in the metallicity distribution cannot be adequately explained by a single intrinsic Gaussian component convolved with the observational uncertainties.

We further tested a Gaussian-plus-power-law functional form for the joint age--metallicity distribution (Appendix Figure~\ref{fig1-app-splash}(a--d)). Within this framework, no restriction beyond the normalizability requirement was imposed on the sign of the power-law slope. The best-fitting solution does not produce an extended metal-poor tail. Instead, the power-law component rises toward lower metallicity and concentrates probability in a relatively narrow feature near $[\mathrm{Fe/H}]\simeq-0.9$.
The Gaussian-plus-power-law model yields $\mathrm{AIC}=4854.08$ and $\mathrm{BIC}=4921.75$, corresponding to $\Delta\mathrm{AIC}=28.91$ and $\Delta\mathrm{BIC}=34.55$ relative to the two-Gaussian model. The information criteria therefore favor the two-Gaussian model, although this preference is less pronounced than that over the single-Gaussian model. The marginal KS tests do not formally reject the Gaussian-plus-power-law model at the conventional 5\% significance level, yielding $D=0.0411$ and $p=0.0918$ for age and $D=0.0387$ and $p=0.1170$ for $[\mathrm{Fe/H}]$. Nevertheless, its agreement with the data is appreciably weaker than that of the two-Gaussian model, which gives $D=0.0168$ and $p=0.9466$ for age and $D=0.0202$ and $p=0.8296$ for $[\mathrm{Fe/H}]$. Thus, although the Gaussian-plus-power-law model remains statistically admissible on the basis of the marginal KS tests alone, its larger KS statistics, lower $p$-values, and higher AIC and BIC values consistently favor the two-Gaussian model.

In addition, we tested the robustness of our results by varying the age--[Fe/H] boundary, the high-$\alpha$/low-$\alpha$ division, and the $L_z$ threshold adopted for the sample selection in Figure~\ref{fig0} within reasonable ranges. These variations do not affect our main conclusions, as demonstrated by the robustness tests presented in Appendix~\ref{app:Splash-Sample-Selection}.

Overall, the model comparison strongly disfavors a single-Gaussian description of the joint age--$[\mathrm{Fe/H}]$ distribution. Both the two-Gaussian and Gaussian-plus-power-law models recover an intrinsic two-component structure consisting of a broad dominant component and a more concentrated, older and metal-poor component. In the fiducial two-Gaussian model, these components produce two distinct but partially overlapping density maxima in the intrinsic age--$[\mathrm{Fe/H}]$ plane, even though their projections do not necessarily appear as two clearly resolved peaks in the age distribution.
We adopt the two-Gaussian model as our fiducial description because it has the lowest AIC and BIC, provides the best agreement
with the observed marginal distributions, and offers a direct decomposition of the intrinsic joint density.


\begin{figure*}[ht!]
\plotone{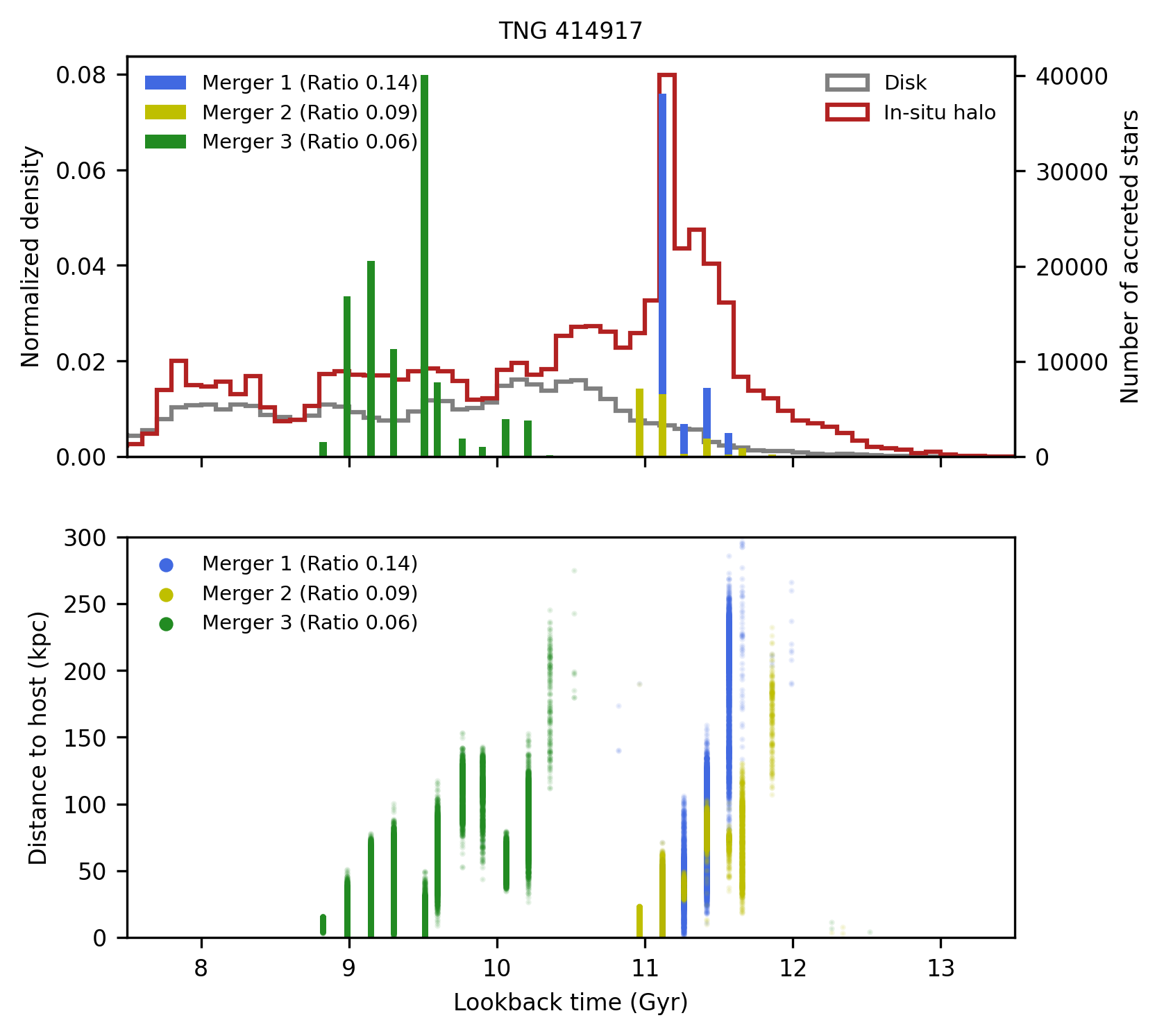}
\caption{Stellar populations in a Milky Way/M31-like galaxy from TNG50 (ID$_{\rm z=0} = 414917$). (Top) Age distributions of disk stars (gray) and in situ halo stars (red), defined as in situ stars with \(L_z \geq 500\ {\rm kpc\,km\,s^{-1}}\) and \(L_z < 500\ {\rm kpc\,km\,s^{-1}}\), respectively. The vertical lines mark the merger times of three satellites with stellar mass ratios of 0.14 (blue), 0.09 (yellow), and 0.06 (green). (Bottom) Galactocentric distances of accreted stars as a function of lookback time.
\label{fig2}}
\end{figure*}

\section{Implications for the early merger history} \label{sec:results-sec3}

The two-component structure inferred in the age--$[\mathrm{Fe/H}]$ distribution suggests that the Splash preserves signatures of more than one phase of star formation and chemical enrichment. Because Splash stars are thought to have formed in the high-$\alpha$ disk before being dynamically heated onto halo-like orbits, this structure may provide a fossil record of temporal substructure in the star formation history of the high-$\alpha$ disk. To examine this connection, Figures~\ref{fig2} and \ref{fig2-app} present two Milky Way analogs from the TNG50 simulation \citep{2019MNRAS.490.3234N,2019MNRAS.490.3196P}. In both systems, the age distribution of the Splash-like population closely traces that of the corresponding high-$\alpha$ disk population, including its principal features. Using a methodology similar to that adopted here, \citet{Sun2026} independently reconstructed the age distribution of the Galactic high-$\alpha$ disk and likewise found evidence for multiple Gaussian components. Taken together, these observational and simulation results support an episodic, rather than entirely smooth, star formation history for the high-$\alpha$ disk, potentially associated with gas-rich mergers and/or discrete episodes of gas accretion.

Moreover, our results imply that the high-$\alpha$ disk experienced a significant dynamical perturbation at a lookback time of $\sim$10.8--11.5~Gyr. The upper bound in lookback time is set by the model-inferred peak of the younger Gaussian component at $\tau\simeq11.5$~Gyr. Because stars must already have formed before they can be dynamically heated, a perturbation at an earlier epoch could not have placed the stars of this younger component onto Splash-like orbits. Conversely, the lower bound is set by the apparent absence of a substantial Splash population associated with the bulk of the high-$\alpha$ disk at $\tau\simeq10.8$~Gyr. Had the perturbation occurred later, it should also have heated a substantial fraction of this dominant disk population, producing a prominent Splash counterpart at these younger ages, contrary to what is observed. The perturbation may have been driven by the accretion and merger of a dwarf galaxy, although the responsible progenitor remains uncertain. One possibility is an early pericentric passage of Gaia--Sausage--Enceladus (GSE; \citealt{2018Natur.563...85H,2018MNRAS.478..611B}), preceding its proposed final major merger phase at a lookback time of $\sim$8--10~Gyr \citep{2018Natur.563...85H,2018MNRAS.478..611B,2021NatAs...5..640M}. 


Recently, \citet{2026arXiv260118896M} inferred a massive accretion event associated with the proposed Low-energy--Kraken--Heracles (LKH) progenitor approximately 1.8~Gyr before GSE, corresponding to a lookback time of $\sim$12.3~Gyr on their adopted absolute age scale. The peak age of $\sim$12.5~Gyr inferred for the older Splash component is broadly consistent with this epoch, suggesting a possible connection with the LKH event. Within the merger-heating scenario, this interaction may have scattered an early population of high-$\alpha$ disk stars onto halo-like orbits, contributing to the older Splash component.

\section{Conclusions} \label{sec:conclusions}

The stellar age distribution of Splash stars provides crucial constraints on the perturbation history of the early Milky Way. In this work, we identify 2,078 candidate Splash stars from a parent sample of more than two million Gaia DR3 subgiants with a typical age precision of $\sim7\%$, representing an approximately tenfold increase over previous samples with comparable age precision. After correcting for the subgiant selection effect, we reconstruct the intrinsic joint age--$[\mathrm{Fe/H}]$ distribution of the Splash population. We find that, on average, Splash stars are older and more metal-poor than the high-$\alpha$ disk, with a selection-corrected age distribution peaking at $\sim12.3$ Gyr, indicating that their formation predates the bulk of the high-$\alpha$ disk.

Forward modeling of the joint age--$[\mathrm{Fe/H}]$ distribution of Splash stars with $[\mathrm{Fe/H}]>-1$ strongly disfavors a single-Gaussian description and reveals an intrinsic two-component structure. Our preferred two-Gaussian model identifies a dominant, relatively younger and more metal-rich component with a peak age of $\sim11.5$ Gyr and $[\mathrm{Fe/H}]\simeq-0.59$, together with an older, more metal-poor component peaking at $\sim12.5$ Gyr and $[\mathrm{Fe/H}]\simeq-0.86$. A Gaussian-plus-power-law model provides an alternative functional form for the metal-poor population, but likewise recovers a concentrated metal-poor component in addition to the broad primary component. Thus, although the precise functional form of the metal-poor component is not uniquely constrained, the presence of an intrinsic two-component structure in the age--$[\mathrm{Fe/H}]$ distribution is robust. This structure suggests that the early high-$\alpha$ disk experienced multiple phases of star formation and chemical enrichment.



The inferred age structure also provides constraints on the dynamical perturbation history of the early high-$\alpha$ disk. Within the merger-heating scenario, the age distribution of the younger Splash component, compared with that of the high-$\alpha$ disk, provides evidence for a substantial dynamical perturbation at a lookback time of approximately 10.8--11.5~Gyr. This perturbation may have been associated with an early passage of Gaia--Sausage--Enceladus (GSE). The older Splash component, with a peak age of $\sim12.5$~Gyr, may preserve signatures of an earlier interaction, possibly associated with the proposed Low-energy--Kraken--Heracles (LKH) event.

\begin{acknowledgments}
We acknowledge financial support from the NSFC through grant Nos. 12503025, 2022000083, 12588202 and the National Key R\&D Program of China grant No. 2022YFF0504200. 
S.B. acknowledges financial support from the National Key R\&D Program of China No. 2024YFA1611901.
This work is based on data acquired through the Guoshoujing Telescope. Guoshoujing Telescope (the Large Sky Area Multi-Object Fiber Spectroscopic Telescope; LAMOST) is a National Major Scientific Project built by the Chinese Academy of Sciences. Funding for the project has been provided by the National Development and Reform Commission. LAMOST is operated and managed by the National Astronomical Observatories, Chinese Academy of Sciences.
This work has made use of data from the European Space Agency (ESA) mission Gaia (\url{https://www.cosmos.esa.int/gaia}), processed by the Gaia Data Processing and Analysis Consortium (DPAC, \url{https://www.cosmos.esa.int/web/gaia/dpac/consortium}). Funding for the DPAC has been provided by national institutions, in particular the institutions participating in the Gaia Multilateral Agreement.

\end{acknowledgments}

\appendix


\clearpage

\setcounter{table}{0}
\renewcommand{\thetable}{A\arabic{table}}


\renewcommand{\thefigure}{A\arabic{figure}}
\setcounter{figure}{0}

\section{Age and Metallicity Distributions With Selection-Function Correction}
\label{app:Subgiant selection}

\begin{figure*}[ht!]
\plotone{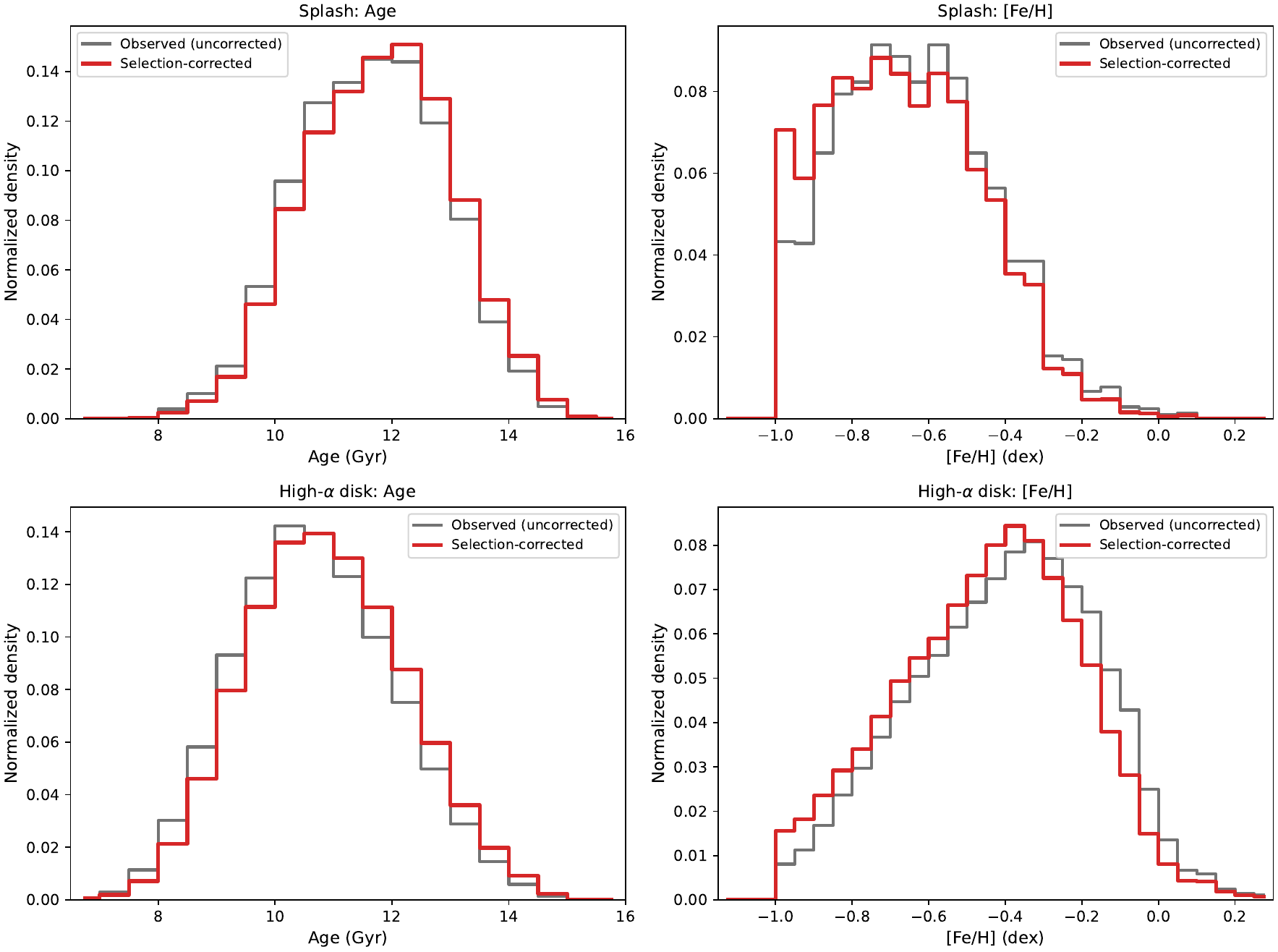}
\caption{Age and metallicity distributions before and after correction for selection effects. The normalized probability-density distributions of age (left panels) and $\mathrm{[Fe/H]}$ (right panels) are shown for the Splash (top panels) and high-$\alpha$ disk (bottom panels) samples. The gray and red histograms represent the observed distributions before correction and the selection-corrected distributions, respectively. The corrected distributions are calculated using normalized inverse-selection weights, $\omega_i \propto 1/s_i$, where $s_i=s(\tau_i,[\mathrm{Fe/H}]_i,[\alpha/\mathrm{Fe}]_i)$. Each histogram is independently normalized to unit integral.
\label{app-selection}}
\end{figure*}

\FloatBarrier

\section{ALTERNATIVE MODELING WITH A GAUSSIAN PLUS POWER-LAW METALLICITY DISTRIBUTION}
\label{app:gauss_powerlaw}

\renewcommand{\thefigure}{B\arabic{figure}}
\setcounter{figure}{0}

\begin{figure*}[ht!]
\centering
\includegraphics[width=0.8\textwidth]{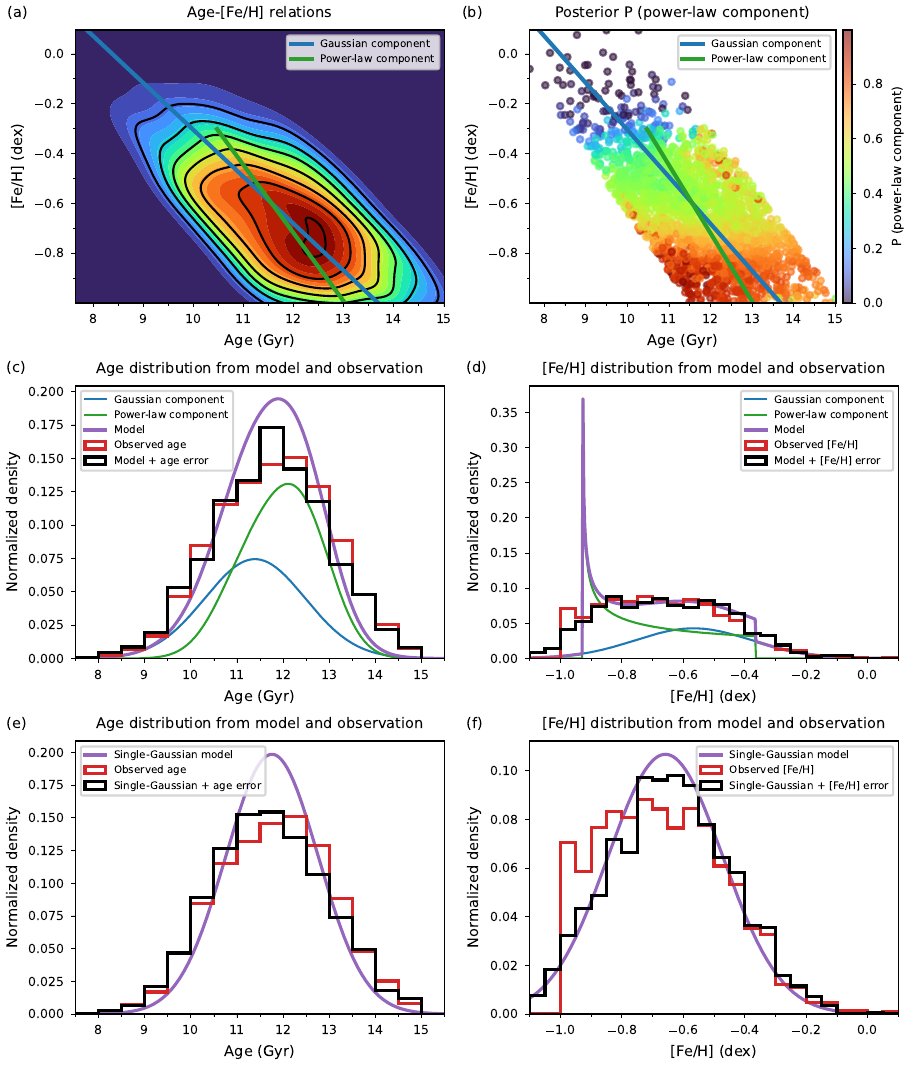}
\caption{Gaussian-plus-power-law and single-Gaussian modeling of the
age--metallicity ($[\mathrm{Fe/H}]$) distribution of the Splash
population. Panel~(a) shows the age--$[\mathrm{Fe/H}]$ distribution as
a selection-weighted two-dimensional KDE, with colors indicating
density and black contours marking iso-density levels. Blue and green
lines show the best-fitting relations of the Gaussian and power-law
components, respectively. The shaded regions indicate the pointwise
16th--84th percentile bootstrap intervals in $[\mathrm{Fe/H}]$ at
fixed age. The bootstrap intervals are narrow over most of the fitted range and
therefore largely overlap the best-fitting relations. The power-law relation and its band are clipped to the
best-fitting $[\mathrm{Fe/H}]$ range. Panel~(b) shows the posterior
probability that each star belongs to the power-law component, with
the same relations and uncertainty bands as in panel~(a).
The second row, panels~(c) and (d), compares the selection-corrected age
and $[\mathrm{Fe/H}]$ distributions with the Gaussian-plus-power-law
model. Blue and green curves show the individual components, and the
purple curve shows their sum. The third row, panels~(e) and (f), shows
the corresponding comparison for the single-Gaussian model. In both
rows, red histograms show the observed distributions and black
histograms show mock distributions after convolution with
observational uncertainties.
\label{fig1-app-splash}}
\end{figure*}

As a robustness test for the intrinsic age–metallicity distribution of the Splash, we considered an alternative model consisting of a Gaussian metallicity component and a truncated power-law metallicity component. This allows the metal-poor part of the distribution to be flexibly described without imposing Gaussianity.

The Gaussian component is modeled in metallicity as
\begin{linenomath*}
\begin{equation}
p_1([\mathrm{Fe/H}]) \sim 
\mathcal{N}(\mu_{{\rm [Fe/H]},1}, \sigma_{{\rm [Fe/H]},1}^2),
\end{equation}
\end{linenomath*}
whereas the truncated power-law component is described as
\begin{linenomath*}
\begin{equation}
p_2([\mathrm{Fe/H}]) \propto 
\left([\mathrm{Fe/H}]-[\mathrm{Fe/H}]_{\min}\right)^\alpha,
\quad 
[\mathrm{Fe/H}]_{\min} \le [\mathrm{Fe/H}] \le [\mathrm{Fe/H}]_{\max},
\quad \alpha > -1 .
\end{equation}
\end{linenomath*}

For both components, the stellar age at fixed latent metallicity is described by a linear relation with intrinsic scatter,
\begin{linenomath*}
\begin{equation}
\tau \mid [\mathrm{Fe/H}],k \sim 
\mathcal{N}\left(a_k[\mathrm{Fe/H}] + b_k,\ \sigma_{\tau,k}^2\right).
\end{equation}
\end{linenomath*}

The total likelihood of a star $i$ with observed age and metallicity 
$\mathbf{x}_i=(\tau_{{\rm obs},i},[\mathrm{Fe/H}]_{{\rm obs},i})$ 
is computed by marginalizing over the latent metallicity. The mixture probability is
\begin{linenomath*}
\begin{equation}
p_i=w\,p_{1,i}+(1-w)p_{2,i},
\qquad
\ln\mathcal{L}_{\rm w}
=
\sum_i\omega_i\ln p_i ,
\label{eq:weighted_powerlaw_likelihood}
\end{equation}
\end{linenomath*}
where $w$ is the weight of the Gaussian component. The metallicity integral is evaluated numerically on a grid extending beyond the observed range, and the intrinsic priors are properly normalized. Multiple initializations are used to reduce sensitivity to local maxima. This formulation preserves the conditional age--metallicity structure while relaxing the Gaussian assumption for the metallicity distribution of the second component.

\FloatBarrier

\section{Robustness Tests of the Splash Sample Selection}
\label{app:Splash-Sample-Selection}

\renewcommand{\thefigure}{C\arabic{figure}}
\setcounter{figure}{0}

\setcounter{table}{0}
\renewcommand{\thetable}{C\arabic{table}}

\begin{figure*}[ht!]
\centering
\includegraphics[width=0.9\textwidth]{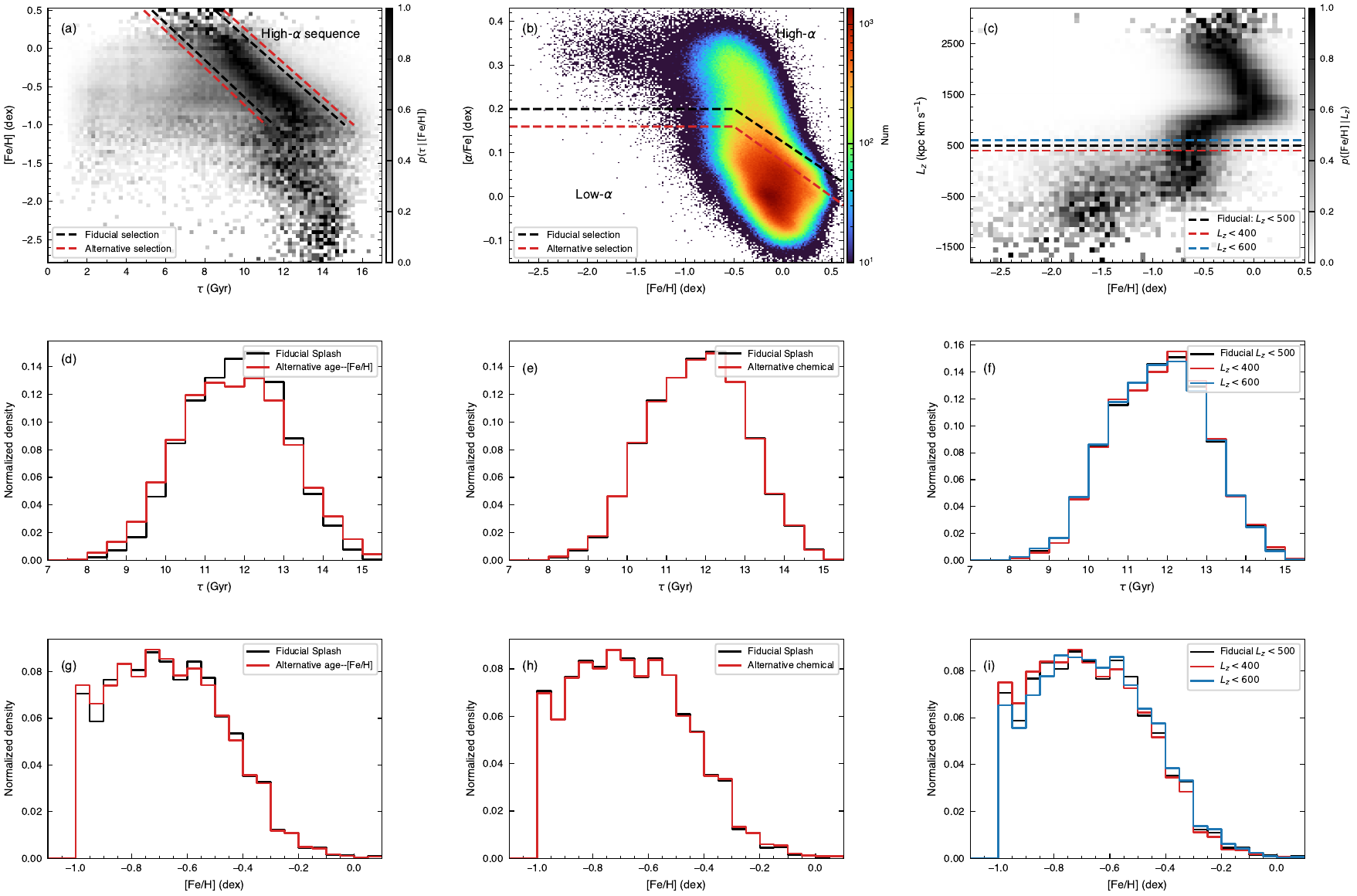}
\caption{Robustness tests of the selection of the Splash population.
(a) Conditional probability distribution of age at a given [Fe/H], $p(\tau \mid \mathrm{[Fe/H]})$, for the high-$\alpha$ disk stars, normalized to the peak value at each [Fe/H] bin. The black and red dashed lines denote the fiducial and broadened age--[Fe/H] selection boundaries, respectively.
(b) Number-density distribution of sample stars in the [Fe/H]--[$\alpha$/Fe] plane. The black and red dashed lines show the fiducial and alternative chemical criteria, respectively, for separating high-$\alpha$ and low-$\alpha$ stars.
(c) Row-normalized stellar density of sample stars in the [Fe/H]--$L_z$ plane, corresponding to $p(\mathrm{[Fe/H]}\mid L_z)$. The horizontal dashed lines indicate the fiducial $L_z<500~{\rm kpc~km~s^{-1}}$ cut (black) and the alternative cuts of $L_z<400$ (red) and $L_z<600~{\rm kpc~km~s^{-1}}$ (blue).
Panels (d)--(f) show the selection-corrected normalized age distributions, while panels (g)--(i) show the selection-corrected normalized metallicity distributions, of the resulting Splash samples. In panels (d, g), the red curves correspond to the broader age--[Fe/H] selection; in panels (e, h), they correspond to the alternative chemical criterion; and in panels (f, i), they show the samples selected with $L_z<400~\mathrm{kpc\,km\,s^{-1}}$ and $L_z<600~\mathrm{kpc\,km\,s^{-1}}$. The black curves show the fiducial Splash sample.
\label{fig:robustness-tests}}
\end{figure*}

\begin{figure*}[ht!]
\centering
\plotone{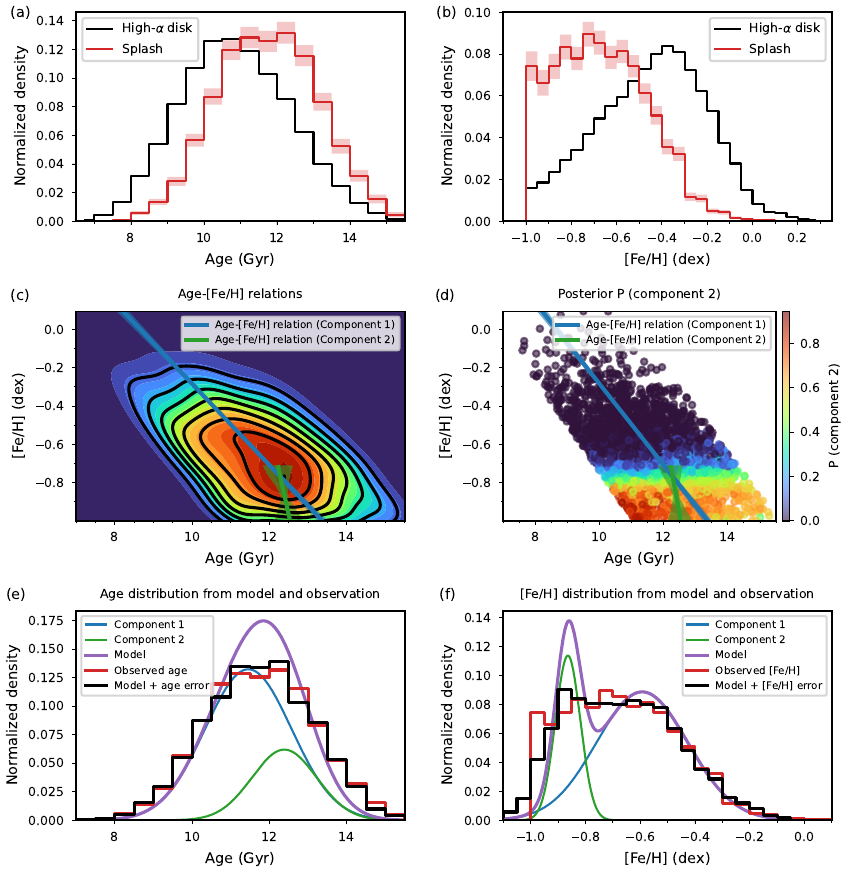}
\caption{Same as Figure~\ref{fig1}, but for the Splash sample selected using the broadened age--$[\mathrm{Fe/H}]$ boundary shown in Figure~\ref{fig:robustness-tests}(a), where the fiducial selection region is expanded by 0.2 dex along the $[\mathrm{Fe/H}]$ direction.
\label{fig:robustness-tests-na-sample}}
\end{figure*}

\begin{figure*}[ht!]
\centering
\plotone{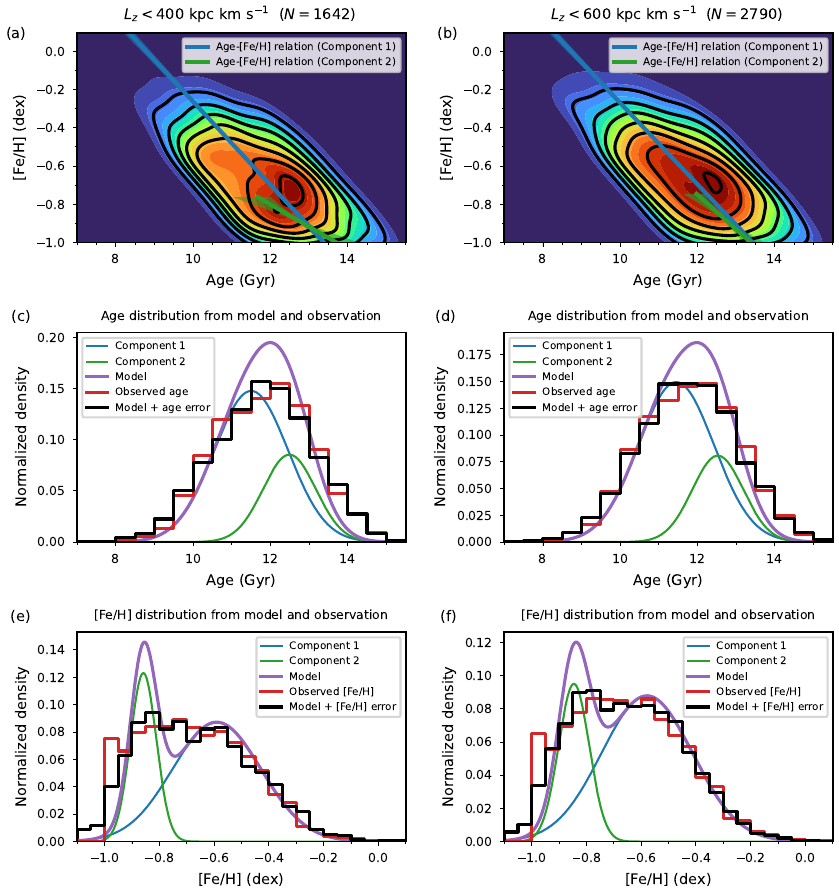}
\caption{Two-Gaussian decomposition of the Splash samples selected with the alternative angular-momentum cuts defined in Figure~\ref{fig:robustness-tests}(c). The left column corresponds to $L_z<400\ {\rm kpc\,km\,s^{-1}}$, and the right column to $L_z<600\ {\rm kpc\,km\,s^{-1}}$. Panels (a) and (b) show the selection-corrected age--$[\mathrm{Fe/H}]$ distributions together with the best-fitting age--$[\mathrm{Fe/H}]$ relations of the two Gaussian components. Panels (c) and (d) show the observed and model age distributions, while panels (e) and (f) show the corresponding $[\mathrm{Fe/H}]$ distributions. All other plotting conventions are the same as in Figure~\ref{fig1}.
\label{fig:robustness-tests-lz-400}}
\end{figure*}

To assess the robustness of our results against the adopted selection criteria, we performed a series of alternative selections for the Splash population. The corresponding selection boundaries and the resulting age and metallicity distributions are shown in Figure~\ref{fig:robustness-tests}. The sample sizes and best-fitting parameters of the two-Gaussian models for these alternative selections are summarized in Table~\ref{tab:robustness-two-component}.


First, we tested the sensitivity of the Splash selection to the adopted age--[Fe/H] boundary. As shown in Figure~\ref{fig:robustness-tests}(a), we broadened the fiducial selection region by extending its width by 0.2 dex in the [Fe/H] direction, resulting in a sample of 2,427 Splash stars. The resulting age and [Fe/H] distributions are shown by the red histograms in Figures~\ref{fig:robustness-tests}(d) and \ref{fig:robustness-tests}(g), respectively. Compared with the fiducial Splash sample adopted in the main text, the age distribution of this enlarged sample is somewhat broader and less sharply peaked, whereas its [Fe/H] distribution remains qualitatively similar. The corresponding two-Gaussian fit is presented in Figure~\ref{fig:robustness-tests-na-sample}. As shown there, the best-fitting model again identifies a metal-poor Gaussian component, with peaks at an age of approximately 12.4~Gyr and [Fe/H] $\simeq -0.86$. These values are consistent with those inferred for the fiducial Splash sample using the two-Gaussian model. This test therefore demonstrates that reasonable variations in the adopted age--[Fe/H] selection boundary do not alter our main conclusion that the intrinsic age--[Fe/H] distribution of the Splash is characterized by two Gaussian components.


Second, we adopted an alternative empirical criterion for separating high- and low-$\alpha$ stars:
\begin{linenomath*}
\[
\left\{
\begin{aligned}
& [\alpha/\mathrm{Fe}] > 0.16, \quad \mathrm{if}\ [\mathrm{Fe/H}] < -0.5,\\
& [\alpha/\mathrm{Fe}] > -0.16[\mathrm{Fe/H}] + 0.08, \quad \mathrm{if}\ [\mathrm{Fe/H}] \geq -0.5.
\end{aligned}
\right.
\]
\end{linenomath*}
The fiducial and alternative chemical boundaries are compared in Figure~\ref{fig:robustness-tests}(b). The resulting Splash population (2,111 stars) has nearly unchanged age and [Fe/H] distributions (Figure~\ref{fig:robustness-tests}(e, h)). This demonstrates that the inferred properties of Splash stars are insensitive to the particular empirical criterion used to define the high-$\alpha$ stars.


Finally, we examined the sensitivity of the Splash selection to the adopted angular-momentum threshold. In addition to the fiducial criterion of $L_z<500~{\rm kpc~km~s^{-1}}$, we considered alternative cuts of $L_z<400$ and $L_z<600~{\rm kpc~km~s^{-1}}$, as illustrated in Figure~\ref{fig:robustness-tests}(c), resulting in samples of 1,642 and 2,790 Splash stars, respectively.
The resulting age and [Fe/H] distributions differ only modestly from those of the fiducial Splash sample adopted in the main text (Figures~\ref{fig:robustness-tests}(f) and \ref{fig:robustness-tests}(i)). The corresponding two-Gaussian fits are shown in Figure~\ref{fig:robustness-tests-lz-400}. In both cases, the best-fitting model identifies a metal-poor component with an age peak of approximately 12.5~Gyr and [Fe/H] peaks of $\simeq-0.86$ and $\simeq-0.85$ for the $L_z<400$ and $L_z<600~{\rm kpc~km~s^{-1}}$ samples, respectively. These results are consistent with those obtained for the fiducial sample, indicating that reasonable variations in the adopted $L_z$ threshold do not affect our main conclusions.


\begin{table*}[ht!]
\centering
\caption{Best-fitting parameters of the two-Gaussian decomposition
for the alternative Splash selections used in the robustness tests.
$N$ is the number of stars in each sample, and $w_k$ is the
selection-corrected mixture fraction of component $k$.
$\mu_{\tau,k}$ and $\mu_{{\rm [Fe/H]},k}$ denote the peak age and
metallicity of each Gaussian component, respectively.
The age--metallicity relation is described by
$\tau=a_k[\mathrm{Fe/H}]+b_k$.
$\delta_{\tau,k}$ is the intrinsic age dispersion at fixed
$[\mathrm{Fe/H}]$, and $\delta_{{\rm [Fe/H]},k}$ is the intrinsic
metallicity dispersion.}
\label{tab:robustness-two-component}

\small
\setlength{\tabcolsep}{3.5pt}
\begin{tabular*}{\textwidth}{@{\extracolsep{\fill}}lccccccccc}
\hline
Selection & $N$ & Comp. & $w_k$
& $\mu_{\tau,k}$
& $\mu_{{\rm [Fe/H]},k}$
& $a_k$
& $b_k$
& $\delta_{\tau,k}$
& $\delta_{{\rm [Fe/H]},k}$ \\
& & & 
& (Gyr)
& (dex)
& (Gyr dex$^{-1}$)
& (Gyr)
& (Gyr)
& (dex) \\
\hline

Broadened age--$[\mathrm{Fe/H}]$
& 2427 & 1 & 0.74
& 11.45 & $-0.59$ & $-4.71$ & 8.66 & 0.79 & 0.167 \\

& & 2 & 0.26
& 12.39 & $-0.86$ & $-1.11$ & 11.43 & 0.84 & 0.046 \\

\hline

$L_z<400~{\rm kpc\,km\,s^{-1}}$
& 1642 & 1 & 0.71
& 11.50 & $-0.59$ & $-4.60$ & 8.79 & 0.60 & 0.162 \\

& & 2 & 0.29
& 12.49 & $-0.86$ & $-8.57$ & 5.13 & 0.55 & 0.047 \\

\hline

$L_z<600~{\rm kpc\,km\,s^{-1}}$
& 2790 & 1 & 0.73
& 11.45 & $-0.58$ & $-4.70$ & 8.73 & 0.59 & 0.166 \\

& & 2 & 0.27
& 12.53 & $-0.85$ & $-5.52$ & 7.86 & 0.58 & 0.056 \\

\hline
\end{tabular*}
\end{table*}


\FloatBarrier

\section{Milky Way/M31-like galaxies from TNG50}
\label{app:tng50-app}

\renewcommand{\thefigure}{D\arabic{figure}}
\setcounter{figure}{0}

\begin{figure*}[ht!]
\plotone{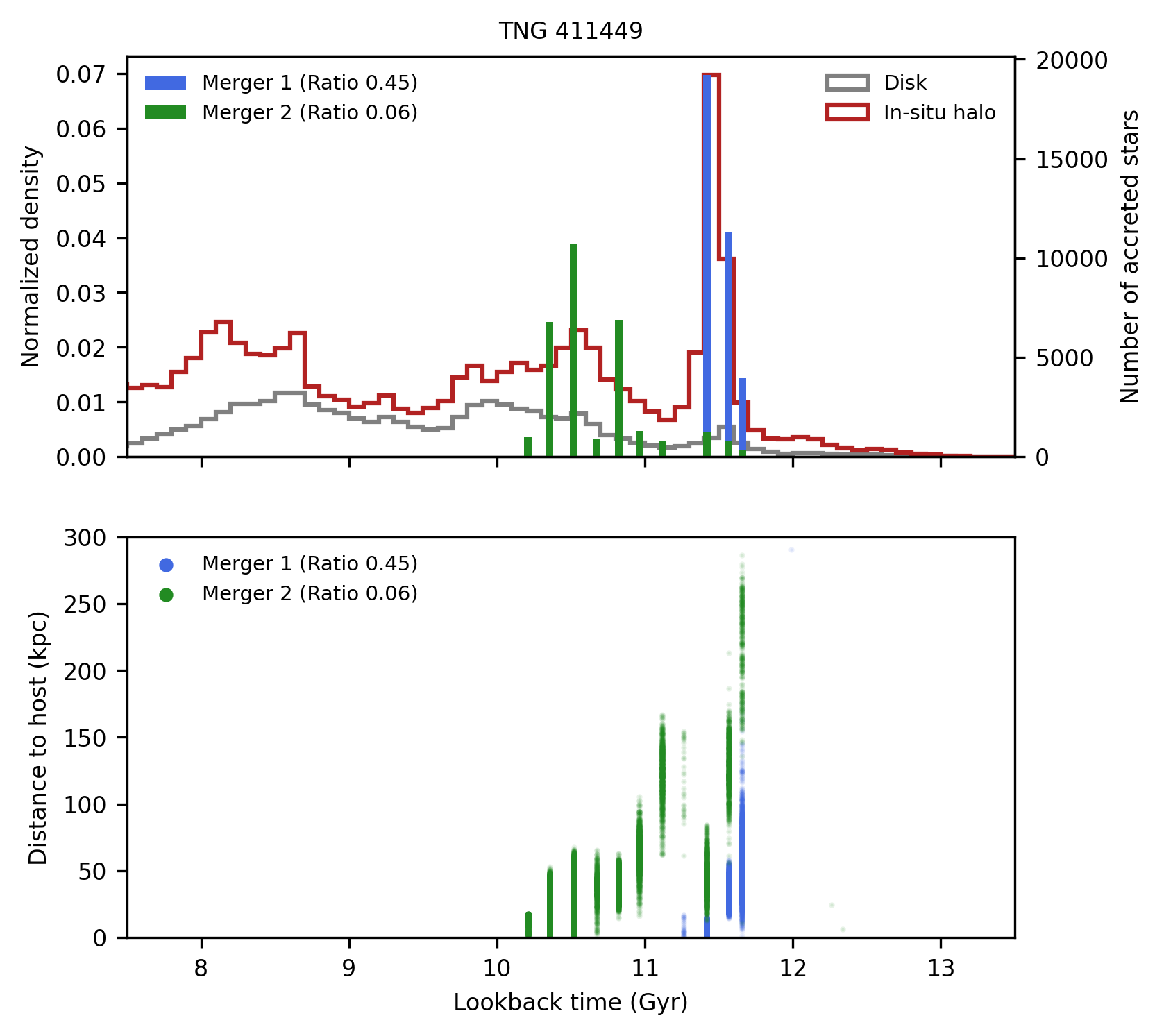}
\caption{Stellar populations in a Milky Way/M31-like galaxy from TNG50 (ID$_{\rm z=0} = 411449$). (Top) Age distributions of disk stars (gray) and in situ halo stars (red), defined as in situ stars with \(L_z \geq 500\ {\rm kpc\,km\,s^{-1}}\) and \(L_z < 500\ {\rm kpc\,km\,s^{-1}}\), respectively. The vertical lines mark the merger times of two satellites with stellar mass ratios of 0.45 (blue) and 0.06 (green). (Bottom) Galactocentric distances of accreted stars as a function of lookback time.
\label{fig2-app}}
\end{figure*}


To place our findings in a theoretical context, we examine the 198 Milky Way– and Andromeda–like (MW/M31-like) galaxies in the TNG50 simulation listed on the TNG project website\footnote{\url{https://www.tng-project.org/data/milkyway+andromeda/}} \citep{2019MNRAS.490.3234N,2019MNRAS.490.3196P}. From this sample, we select two analogs, TNG-414917 and TNG-411449, whose distributions in the $[\mathrm{Fe/H}]$–$L_z$ plane resemble that observed for the Milky Way (Figure~\ref{fig0}(d)) and whose early merger histories are broadly comparable to that inferred for the Galaxy. As shown in Figure~\ref{fig2} and Figure~\ref{fig2-app}, both systems underwent multiple mergers at lookback times greater than 8~Gyr. In each case, the most massive merger occurred more than 11~Gyr ago and contributed most efficiently to the formation of the in situ halo.




\clearpage

\bibliography{bibliography}{}
\bibliographystyle{aasjournalv7}



\end{document}